\documentclass[prl,twocolumn,showpacs,superscriptaddress]{revtex4-1}
\usepackage{epsfig}
\usepackage{graphicx}
\usepackage{dcolumn}
\usepackage{bm}
\usepackage{overpic}
\usepackage{subfigure}
\usepackage{float}
\usepackage{color}
\usepackage{amsmath}
\usepackage{mathcomp}
\usepackage{appendix}
\usepackage{changes}
\usepackage{adjustbox}
\usepackage[absolute,overlay]{textpos}

\usepackage[linktoc=all]{hyperref}

\newcommand{\gevcc}{\ensuremath{{\mathrm{GeV/}c^2}~}}
\newcommand{\gevgevcccc}{\ensuremath{{\mathrm{GeV^2/}c^4}~}}
\providecommand{\papersection[1]}{\paragraph{#1}}

\begin{document}
\title{ \boldmath Study of the $f_{0}(980)$ and $f_{0}(500)$ Scalar Mesons through the Decay $D_{s}^{+} \to \pi^{+} \pi^{-} e^{+} \nu_{e}$ }
\author{
M. Ablikim \emph{et al.$^{*}$}\\
(BESIII Collaboration)\\
}
\date{8, April, 2024}

\begin{abstract}
Using $e^+e^-$ collision data corresponding to an integrated luminosity of 7.33~${\rm fb^{-1}}$ recorded by the BESIII detector
at center-of-mass energies between 4.128 and 4.226~${\rm GeV}$,
we present an analysis of the decay $D_{s}^{+} \to \pi^{+}\pi^{-} e^{+}\nu_{e}$, where the $D_s^+$ is produced via the process $e^+e^- \to D_{s}^{*\pm}D_{s}^{\mp}$.
We observe the $f_{0}(980)$ in the $\pi^+\pi^-$ system and the branching fraction of 
the decay $D_{s}^{+} \to f_{0}(980)e^{+}\nu_{e}$ with $f_0(980)\to\pi^+\pi^-$ measured to be $(1.72 \pm 0.13_{\rm stat} \pm 0.10_{\rm syst}) \times10^{-3}$,
where the uncertainties are statistical and systematic, respectively.
The dynamics of the $D_{s}^{+} \to f_{0}(980)e^{+}\nu_{e}$ decay are studied with the simple pole parametrization of the hadronic form factor
and the Flatt\'e formula describing the $f_0(980)$ in the differential decay rate,
and the product of the form factor $f^{f_0}_{+}(0)$ and the $c\to s$ Cabibbo-Kobayashi-Maskawa matrix element $|V_{cs}|$
is determined for the first time to be $f^{f_0}_+(0)|V_{cs}|=0.504\pm0.017_{\rm stat}\pm0.035_{\rm syst}$.
Furthermore, the decay $D^+_s\to f_0(500) e^+\nu_e$ is searched for the first time but no signal is found.
The upper limit on the branching fraction of $D^+_s\to f_0(500) e^+\nu_e,~f_0(500)\to\pi^+\pi^-$ decay
is set to be $3.3 \times 10^{-4}$ at 90\% confidence level.
\end{abstract}
\maketitle

\begin{textblock*}{8cm}(\oddsidemargin+2.5cm, \paperheight-2.0cm)
  \footnotesize 
  \raggedright  
  \hrule width 3cm height 0.4pt \vspace{0.5em}
  \textbf{\ \ \ \ \ *Full author list given at the end of the Letter.} \\
\end{textblock*}

\papersection{Introduction}
Quantum Chromodynamics (QCD), the fundamental theory of the strong interaction, has been established for almost half a century.
However, there are some features that still need to be understood,
such as quark confinement and dynamics in the nonperturbative regime. 
The light scalar mesons $f_0(500)$, $f_0(980)$, and $a_0(980)$ play a crucial role in the dynamics of the spontaneous breaking of QCD chiral symmetry and in the origin of pseudoscalar meson masses~\cite{Pelaez2016,weiwang2010},
and consequently can be used to probe the confinement of quarks~\cite{Jaffe1977}.
Furthermore, our understanding of the nature of light hadrons is still poor since QCD is nonperturbative in the low-energy region.
Investigating the structure of the light scalar mesons provides key input to these issues.
In spite of the striking success of the constituent quark model,
the nontrivial quark structure of these mesons has remained controversial for many years~\cite{pdg}.
Their mass ordering cannot be explained by a $q\bar{q}$ configuration in the naive quark model,
leaving open the possibility that they are mixtures of $q\bar{q}$ 
states~\cite{Jaffe1977,qqbar1,qqbar2,QCDSR1,QCDSR2,LCSR1,LCSR2,LF-RQM,Soni2020,Cheng,RMWang2023,YKHsiao2023}.
Other interpretations are diquark-antidiquark states (tetraquark)~\cite{tetraquark} 
and meson-meson bound states (molecule)~\cite{molecule}.
Therefore, more conclusive experimental measurements of these scalar states are highly desired.

Since the leptons and hadrons in the final state interact only weakly with each other,
semileptonic decays of charm mesons provide a unique and clean platform to probe
the constituent $q\bar{q}$ components in the wave functions of light scalar states~\cite{Achasov2012}. 
Here, only the spectator light quarks are related to the formation of these states and the quark flavor content can be specified through Cabibbo-favored and -suppressed processes~\cite{Oset2015}.
Additionally, the dynamics of the semileptonic charmed meson decays can be studied by measuring the hadronic form factor 
that describes the strong interaction between the final-state quarks, including all the nonperturbative effects.
This provides an excellent opportunity to test the different theoretical methods of solving the QCD nonperturbative problem.
Since the form factors and branching fractions (BFs) of the semileptonic charmed meson decays are highly sensitive to the internal structure of light scalar states, 
studies of the dynamics of these decays are also important to understand their nature~\cite{Soni2020}.

In previous studies, the BESIII Collaboration has reported measurements of
the decays $D^{0(+)}\to a_{0}(980)^{-(0)}e^{+}\nu_{e}$~\cite{bes3-D2a0enu}, $D^{+}\to f_{0}(500) e^{+}\nu_{e}$~\cite{bes3-D2f0enu},
and $D_{s}^{+}\to f_{0}(980)e^{+}\nu_{e}$ with $f_{0}(980)\to \pi^{0}\pi^{0}$~\cite{bes3-Ds2f0enu},
and searches of the decays $D^{+}\to f_{0}(980) e^{+}\nu_{e}$~\cite{bes3-D2f0enu}, $D_{s}^{+}\to a_{0}(980)^{0}e^{+}\nu_{e}$~\cite{bes3-Ds2a0enu} and $D_{s}^{+}\to f_{0}(500)e^{+}\nu_{e}$ with $f_{0}(500)\to \pi^{0}\pi^{0}$~\cite{bes3-Ds2f0enu}.
With negligible contamination from the $D_{s}^{+} \to \rho^{0} e^+ \nu_{e}$ channel, the decay $D_{s}^+\to \pi^{+}\pi^{-}e^{+}\nu_{e}$
enables us to study the structure of $f_{0}(980)$ in a clean environment.
Previously, only the CLEO Collaboration measured the BF of the decay
$D_{s}^{+}\to f_{0}(980)e^{+}\nu_{e}$ with $f_{0}(980)\to \pi^{+}\pi^{-}$~\cite{cleo2009-Ds2semilep,cleo2009-Ds2f0enu,cleo2015-Ds2semilep} with data taken at a center-of-mass (CM) energy ($E_{\rm CM}$) near 4.170 GeV.
With a data sample more than 10 times larger, 
we report a significantly improved measurement of the BF and the first measurement of the transition form factor of the decay $D_{s}^{+}\to f_{0}(980)e^{+}\nu_{e}$ with $f_{0}(980)\to \pi^{+}\pi^{-}$, 
and the first search of the decay $D_{s}^{+}\to f_{0}(500)e^{+}\nu_{e}$ with $f_{0}(500)\to\pi^{+}\pi^{-}$.  
The obtained results are important tests of theoretical predictions based on different models~\cite{qqbar2,QCDSR1,QCDSR2,LCSR1,LCSR2,LF-RQM,Soni2020,RMWang2023,YKHsiao2023}.
Throughout this Letter, charge-conjugate channels are implied.

\papersection{BESIII experiment and data samples}
For the BF measurements of semileptonic decays, we use the same tag technique of Refs~\cite{bes3-Ds2Kenu,bes3-Ds2a0enu,bes3-Ds2f0enu} with additional detail contained in Appendix~A.
Our measurements are performed based on $e^+e^-$ collision data corresponding to an integrated luminosity of 7.33~fb$^{-1}$
collected with the BESIII detector at $E_{\rm CM}=4.128 - 4.226$~GeV~\cite{material}.
Details about the BESIII detector design and performance are provided in Refs.~\cite{BESIII1,BESIII2,BESIII3}. 

Simulated data samples produced with a {\sc geant4}-based~\cite{GEANT4} Monte Carlo (MC) package,
which includes the geometric description of the BESIII detector~\cite{BESIII4} and the detector response,
are used to determine detection efficiencies and to estimate backgrounds.
The simulation models the beam energy spread and initial state radiation in the $e^+e^-$ annihilations with the generator {\sc kkmc}~\cite{KKMC}.
The inclusive MC sample includes the production of open-charm processes, the initial state radiation production of vector charmonium(-like) states,
and the continuum processes incorporated in {\sc kkmc}~\cite{KKMC}.
All particle decays are modeled with {\sc evtgen}~\cite{EVTGEN} using BFs either taken from the Particle Data Group~\cite{pdg},
when available, or otherwise estimated with {\sc lundcharm}~\cite{LUNDCHARM}.
Final state radiation from charged final state particles is incorporated using the {\sc photos} package~\cite{PHOTOS}.
The signal detection efficiencies and signal shapes are obtained from the signal MC samples,
in which the $D_{s}^{-}$ decays inclusively to all known decay channels and the signal $D_{s}^{+}$ decays to $\pi^{+}\pi^{-}e^{+}\nu_{e}$
with the $S$-wave contribution simulated according to previous measurements~\cite{bes3-Dp2kpienu,bes3-D2f0enu}.
The amplitudes for the $f_{0}(980)$ is modeled by the Flatt\'e formula with its parameters fixed to the BESII measurement~\cite{bes2f980}.  

\papersection{Event selection}
The tag $D^-_s$ candidates are reconstructed with $K^{\pm}$, $\pi^{\pm}$, $\rho^{-}$, $\rho^{0}$, $\pi^{0}$, $\eta^{(')}$, and $K_{S}^{0}$ mesons in 12 tag modes:
$K^{+}K^{-}\pi^{-}$,
$K^0_{S}K^{-}$,
$\pi^{-}\eta$,
$\pi^{-}\eta^{\prime}_{\pi^{+}\pi^{-}\eta}$,
$K^{+}K^{-}\pi^{-}\pi^{0}$,
$\pi^{+}\pi^{-}\pi^{-}$,
$K^0_{S}K^{+}\pi^-\pi^-$,
$\rho^{-}\eta$,
$\pi^{-}\eta^{\prime}_{\gamma\rho^0}$,
$K^{+}\pi^{-}\pi^{-}$,
$K^{0}_{S}K^{-}\pi^{0}$,
and $K^0_{S}K^{-}\pi^+\pi^-$.
A detailed description of the selection criteria for all tag candidates except $\rho^{-}\eta$
can be found in Ref~\cite{Suntong}. 
The $\rho^-$ candidates are reconstructed from $\pi^-\pi^0$ combinations within an invariant mass interval (0.625, 0.925) GeV/$c^2$.
Requirements on the recoiling mass ${m_{\rm rec}}$ against the tag $D_{s}^{-}$ candidates are applied to the tag candidates in order to identify the process $e^+e^- \to D_{s}^{*\pm}D_{s}^{\mp}$.
If there are multiple candidates for a specific tag mode per charge, 
the one with ${m_{\rm rec}}$ closest to the known $D_{s}^{*\pm}$ mass~\cite{pdg} is chosen.
For each tag mode, the tag yield is extracted from the fit to the tag $D_{s}^{-}$ mass spectrum ($M_{D_{s}^{-}}$).
The signals are modeled with the MC-simulated signal shape convolved with a Gaussian function
to account for the resolution difference between data and MC simulation,
while the combinatorial backgrounds are parameterized with a first-order or second-order Chebyshev polynomial.
For the tag mode $D^-_s\to K^0_SK^-$, the peaking background from $D^-\to K^0_S\pi^-$ decay is described by the MC-simulated shape that is smeared
with the same Gaussian function as used in the signal, with the background yield determined from the fit.
Summing over various tag modes and energy points, 
we obtain the total tag yield $N^{\rm tot}_{\rm tag}=771~101\pm3445$.
For more details about tag candidates, such as selection regions and reconstruction efficiencies, see Ref.~\cite{material}.

After a $D_{s}^{-}$ candidate is identified, we reconstruct the decay $D_{s}^{+}\to \pi^+\pi^- e^{+}\nu_{e}$ recoiling against the tag side,
requiring three charged tracks identified as a $\pi^+\pi^-$ pair with the same selection criteria as on the tag side 
and $e^+$ (opposite sign to the tag $D_{s}^{-}$) following Ref.~\cite{bes3-Dp2omegaenu}.
Using the same kinematic fit method of Refs~\cite{bes3-Ds2Kenu},
we reconstruct the transition photon from the main decay $D_{s}^{*\pm}\to\gamma D_{s}^{\pm}$. 

For the real $D^{*\pm}_sD^{\mp}_s$ events, the square of the recoiling mass (${M^2_{\rm rec}}$) against the transition photon
and the tag $D_{s}^{-}$ is expected to peak at the known $D_{s}^{+}$ mass squared.
To improve the resolution, the decay products of the tag $D^-_s$ are constrained to the known $D^{+}_s$ mass~\cite{pdg}.
We require ${M^2_{\rm rec}}$ to be within (3.78, 4.05) $\gevgevcccc$
to suppress the backgrounds from non-$D_{s}^{*\pm}D_{s}^{\mp}$ processes.
The missing neutrino information is inferred by the missing mass squared which is defined as
\begin{equation}
M^{2}_{\rm miss}=(\bm{p}_{\rm CM}-\bm{p}_{\rm tag}-\bm{p}_{\pi^+}-\bm{p}_{\pi^-}-\bm{p}_{e}-\bm{p}_{\gamma})^2,
\label{eq:MM2}
\end{equation}
where $\bm{p}_{\rm CM}$ is the four-momentum of the $e^+e^-$ center-of-mass system, $\bm{p}_{\rm tag}$ for the tag $D_{s}^{-}$, $\bm{p}_{\pi^+(\pi^-,e)}$ for the semileptonic final state,
and $\bm{p}_{\gamma}$ for the transition photon from the $D_{s}^{*\pm}$ decay.
Here, the measured momenta of the tag $D^-_s$ and the transition photon are corrected with the kinematic fit to improve the resolution.
In order to further reject backgrounds, we require $|M^{2}_{\rm miss}|< {\rm 0.06~GeV^{2}}/c^{4}$.

\begin{figure}[htp]
\begin{center}
\includegraphics[width=0.455\textwidth]{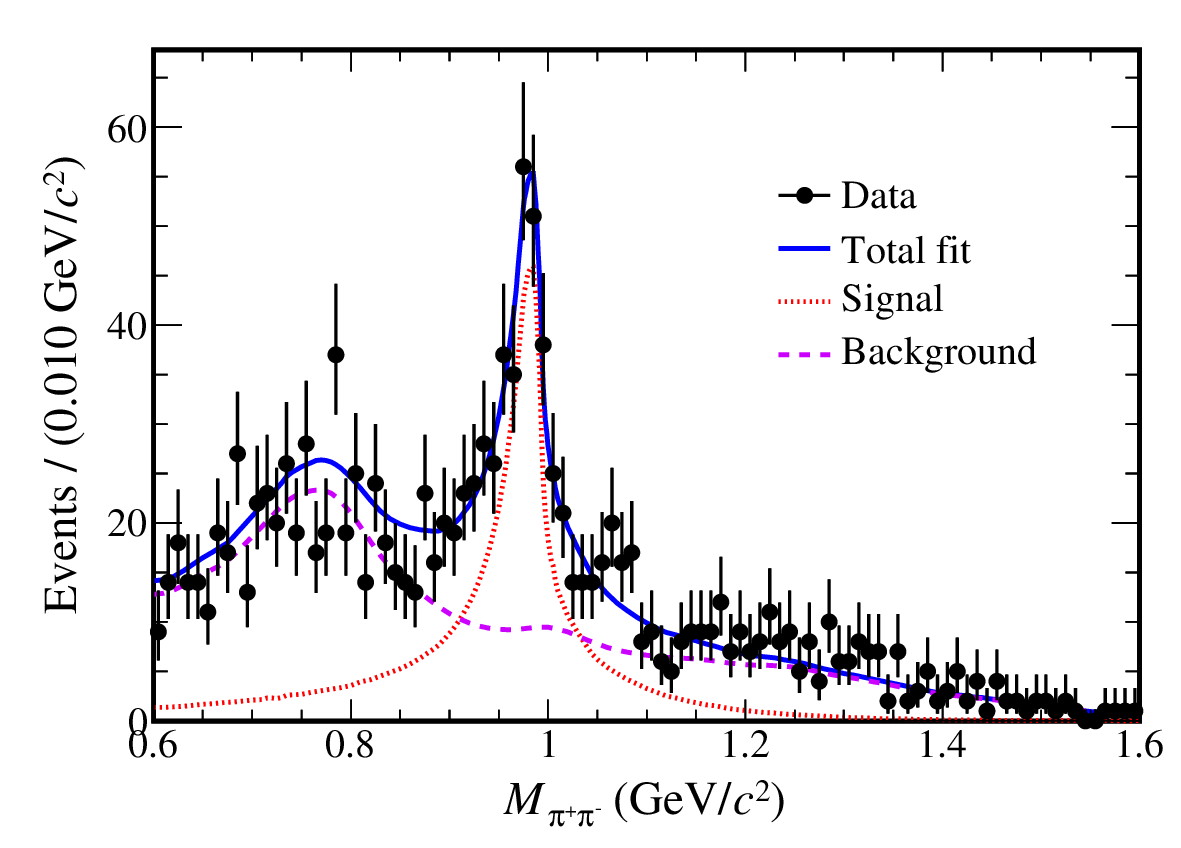}
\put(-82,83){ {\scriptsize $\chi^2/{\rm NDF}=0.7$} }
\caption{ Fit to the $M_{\pi^+\pi^-}$ distribution of the accepted candidates for the decay $D_{s}^{+} \to f_0(980) e^{+}\nu_{e}$.
The points with error bars are data, and the blue line is the total fit.
The red dotted and violet dashed lines are the signal and background shapes, respectively.
}
\label{fit:hadMass-data}
\end{center}
\end{figure}

\papersection{BF measurement}
To study the $f_{0}(980)$, we require the $\pi^+\pi^-$ invariant mass ($M_{\pi^+\pi^-}$, see Fig.~\ref{fit:hadMass-data}) to be within the interval (0.6, 1.6) ${\rm GeV}/c^{2}$.
The nonpeaking background distribution from the inclusive MC sample is verified using events from the data sideband region 
(about 2$\sigma$ away from the signal region of the tag $D_{s}^{-}$ mass and having the same interval as signal region)
of the tag $M_{D_{s}^{-}}$ distribution.
The peak around 0.77~\gevcc is mainly caused by the decay $D^+_s\to\eta^{\prime}(\gamma\pi^+\pi^-)e^+\nu_e$.
We find that the inclusive MC adequately describes the data in this channel.
An unbinned maximum likelihood fit to the $M_{\pi^+\pi^-}$ distribution is performed to extract the signal yield of $D_{s}^{+}\to f_{0}(980)e^{+}\nu_{e},~f_{0}(980)\to \pi^{+}\pi^{-}$ decay.
In the fit, the signal is modeled with an MC-simulated line shape convolved with a Gaussian resolution function,
and the background is described by the inclusive MC shape convolved with the same Gaussian function.
From the fit, which is shown in Fig.~\ref{fit:hadMass-data}, we obtain the total signal yield $N^{\rm tot}_{\rm sig}=439\pm33$.
The goodness of fit $\chi^2/{\rm NDF}$ is 0.7, where ${\rm NDF}$ is the number of degrees of freedom.
Particularly, the other $S$-wave contributions from the $f_0(500)$, $f_0(1370)$ and nonresonance can be ignored since no significant signal is observed.
Using the (35.44$\pm$0.07)\% weighted efficiency provided in Ref.~\cite{material} and the formula in Appendix~A, 
we obtain $\mathcal{B}[D^+_s\to f_0(980)e^{+}\nu_{e}, f_0(980)\to\pi^+\pi^-] = (1.72\pm0.13_{\rm stat}\pm0.10_{\rm syst})\times 10^{-3}$
where the systematic uncertainties are discussed in Appendix~B.
Using the BF ${\cal B}[f_0(980)\to\pi^+\pi^-]=(46\pm6)\%$ assuming the dominant $\pi\pi$ and $KK$ decays~\cite{LHCb-B02Jpipi} 
and the relation ${\cal B}[D^+_s\to f_0(980)e^+\nu_e]=4.22\times 10^{-3}{\rm cos}^2\phi$ with the mixing angle $\phi$ involved in the $q\bar{q}$ mixture picture 
for $f_0(980)$ as ${\rm sin}\phi\frac{1}{\sqrt{2}}(u\bar{u}+d\bar{d})+{\rm cos}\phi s\bar{s}$~\cite{QCDSR2,LF-RQM}, 
we obtain the angle $\phi=(19.7\pm12.8)^{\circ}$ implying that the $s\bar{s}$ component dominates.
 
\begin{figure}[htp]
\begin{center}
\includegraphics[width=0.425\textwidth]{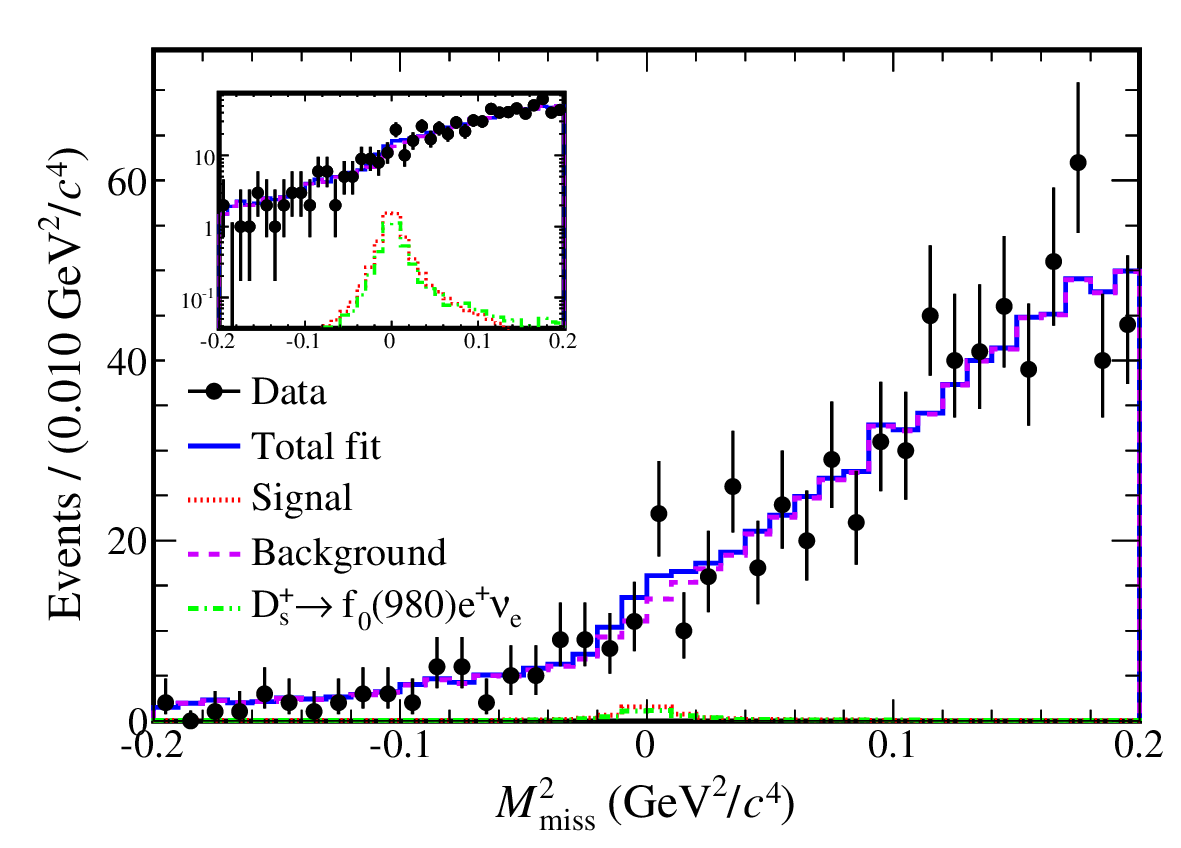}
\put(-148,72){\fontsize{6}{6} { $\chi^2/{\rm NDF}=0.6$} }
\caption{Fit to the $M^{2}_{\rm miss}$ distribution of the accepted candidates for $D_{s}^{+} \to f_{0}(500) e^{+} \nu_{e}$ decay.
The black points with error bars are data, and the blue line is the total fit.
The red dotted and violet dashed lines are the signal and background shapes, respectively.
The green dotted-dashed line is the decay $D_s^+\to f_0(980)e^+\nu_e$.
}
\label{ul-fit}
\end{center}
\end{figure}

We further search for the decay $D_{s}^{+}\to f_{0}(500)e^{+}\nu_{e}$ with $f_{0}(500)\to\pi^{+}\pi^{-}$.
To avoid the background from $D_{s}^{+}\to K_{S}^{0} e^{+}\nu_{e}$ decay and the possible tail of $D_{s}^{+}\to f_{0}(980)e^{+}\nu_{e}$ decay,
we only use the events satisfying $M_{\pi^{+}\pi^{-}} < \mathrm{0.45~GeV}/c^{2}$.
The background yield of $D_{s}^{+}\to f_{0}(980)e^{+}\nu_{e},~f_{0}(980)\to\pi^{+}\pi^{-}$ decay is estimated to be 5.4 based on the foregoing study.
An unbinned maximum likelihood fit to the $M^{2}_{\rm miss}$ distribution of the accepted candidates is performed,
where the signal and background shapes are modeled by the simulated shapes obtained from the signal and inclusive MC samples, respectively.
The fit result is shown in Fig.~\ref{ul-fit}.
Since no significant signal is observed, 
an upper limit on the BF at 90\% confidence level with the (11.98$\pm$0.06)\% weighted efficiency is set to be
$\mathcal{B}[D_{s}^{+}\to f_{0}(500)e^{+}\nu_{e}, f_{0}(500) \to \pi^{+}\pi^{-}]<3.3 \times 10^{-4}$ following Ref~\cite{bes3-Ds2f0enu}.
The related systematic uncertainties are discussed in Appendix~B. 


\papersection{Form factor measurement}
The dynamics of $D_{s}^{+}\to f_{0}(980)e^{+}\nu_{e}$ decay is studied by dividing the semileptonic candidate events into four intervals of $q^2$ 
(four-momentum transfer square of $e^{+}\nu_{e}$).
Using the measured and expected partial decay rates of the $i$th $q^{2}$ interval,
$\Delta\Gamma^{i}_{\rm mea}$ and $\Delta\Gamma^{i}_{\rm exp}$, the form factor is determined by constructing and minimizing a $\chi^2$ as 
\begin{equation}
\chi^{2}=\sum_{ij} (\Delta\Gamma^{i}_{\rm mea} - \Delta\Gamma^{i}_{\rm exp})(C^{-1})_{ij}(\Delta\Gamma^{j}_{\rm mea} - \Delta\Gamma^{j}_{\rm exp}),
\label{eq:chi}
\end{equation}
where $C_{ij}$ is the covariance matrix to consider correlations of $\Delta\Gamma^{i}_{\rm mea}$ among $q^2$ intervals.

The $\Delta\Gamma^{i}_{\rm exp}$ is calculated by integrating the following double differential decay rate~\cite{FF-rate-sq}
\begin{align}
\frac{d^{2}\Gamma(D^+_s\to f_0(980)e^+\nu_e)}{dsdq^2}=&\frac{G^2_F|V_{cs}|^{2}}{192\pi^{4}m_{D^+_s}^{3}}\lambda^{3/2}(m_{D^+_s}^{2},s,q^2) \nonumber \\
& \times |f^{f_0}_+(q^2)|^2 P(s),
\label{eq:rate}
\end{align}
where $s$ is the square of $M_{\pi^+\pi^-}$, $G_F$ is the Fermi constant~\cite{pdg}, 
$|V_{cs}|$ is the Cabibbo-Kobayashi-Maskawa matrix element, $m_{D^+_s}$ is the known $D^+_s$ mass~\cite{pdg},
$\lambda(x,y,z)=x^2+y^2+z^2-2xy-2xz-2yz$,
and $P(s)$ is based on the relativistic Flatt\'e formula~\cite{bes2f980} due to the open $K^+K^-$ channel as follows:
\begin{equation}
P(s)=\frac{g_1\rho_{\pi\pi}}{|m^2_0-s-i(g_1\rho_{\pi\pi} + g_2\rho_{K\bar{K}})|^2}.
\label{eq:Ps}
\end{equation}
Here $m_0$ denotes the $f_0(980)$ mass;
the constants $g_1$ and $g_2$ are the $f_0(980)$ couplings to $\pi^+\pi^-$ and $K^+K^-$ final states, respectively;
and $\rho_{\pi\pi}$ and $\rho_{K\bar{K}}$ are individual phase space factors.
Using the decay widths in the different $q^2$ intervals, the form factor $|f^{f_0}_+(q^2)|$ can be extracted.
In this Letter, the form factor is modeled with the simple pole parametrization~\cite{FF-simple}:
\begin{equation}
f^{f_0}_+(q^2)=\frac{f^{f_0}_+(0)}{1-q^2/M^2_{\rm pole}},
\label{eq:SPD}
\end{equation}
where $f^{f_0}_+(0)$ is the form factor evaluated at $q^2=0$, and the pole mass $M_{\rm pole}={\rm 2.46~GeV}/c^2$~\cite{pdg,ref-Ds1}.

The measured partial decay rate $\Delta\Gamma_{\rm mea}^{i}$ is determined by 
$\Delta\Gamma_{\rm mea}^{i} \equiv \int_{i}\int_{s} \frac{d^{2}\Gamma}{dsdq^{2}}dsdq^{2}=N^i_{\rm pro}/(\tau N^{\rm tot}_{\rm tag}{\cal B}_{\gamma})$,
where ${\cal B}_{\gamma}$ represents the BF of $D_{s}^{*\pm}\to\gamma D_{s}^{\pm}$,
$\tau$ is the $D_{s}^{+}$ meson lifetime~\cite{pdg,lifetime}, and $N_{\rm pro}^{i}$ is the signal yield produced in the $i$th $q^2$ interval,
obtained as $N_{\rm pro}^{i}=\sum_{j=1}^{4} \epsilon^{-1}_{ij} N_{\rm obs}^{j}$.
Here, $N_{\rm obs}^{j}$ is the observed signal yield obtained from a fit to the $M_{\pi^{+}\pi^{-}}$ distribution
in the $j$-th $q^2$ interval, which is carried out in a similar manner as the one described previously for the BF measurement, 
and $\epsilon_{ij}$ is the efficiency matrix determined from the signal MC samples via 
$\epsilon_{ij}=\sum_{k}[(1/N_{\rm tag}^{\rm tot}) \times (N_{\rm rec}^{ij}/N_{\rm gen}^{j})_{k} \times (N_{\rm tag}^{k}/\epsilon_{\rm tag}^{k})]$,
where $N_{\rm rec}^{ij}$ is the signal yield reconstructed in the $i$th $q^2$ interval and generated in the $j$-th $q^2$ interval,
$N_{\rm gen}^{j}$ is the total signal yield generated in the $j$-th $q^2$ interval, and $k$ sums over all tag modes.
The details of the divisions, $N_{\rm obs}^{i}$ and $\Delta\Gamma_{\rm mea}^{i}$ of various $q^2$ intervals are given in Ref.~\cite{material}.

The statistical and systematic covariance matrices are constructed as 
$C_{ij}^{\rm stat}=[1/(\tau N_{\rm tag}^{\rm tot})^{2}]\sum_{\alpha}\epsilon_{i\alpha}^{-1}\epsilon_{j\alpha}^{-1}\sigma^{2}(N^{\alpha}_{\rm obs})$ 
and $C_{ij}^{\rm syst}=\delta(\Delta\Gamma_{\rm mea}^{i})\delta(\Delta\Gamma_{\rm mea}^{j})$, respectively,
where $\sigma(N^{\alpha}_{\rm obs})$ and $\delta(\Delta\Gamma_{\rm mea}^{i})$ are the statistical and systematic uncertainties in the $i$th $q^2$ interval.
The $C_{ij}^{\rm syst}$ is obtained by summing all the covariance matrices for all systematic uncertainties, 
where the systematic uncertainty of $\tau$, 0.8\%~\cite{pdg,lifetime}, is involved besides those in the BF measurement.
The obtained $C_{ij}^{\rm stat}$ and $C_{ij}^{\rm syst}$, the resulting $C_{ij}=C_{ij}^{\rm stat}+C_{ij}^{\rm syst}$, 
and the relevant correlation matrix element $\rho_{ij}$ are shown in Ref.~\cite{material}.

The systematic uncertainty related to $\Delta\Gamma_{\rm mea}^{i}$ is estimated to be 2.6\% by following Ref~\cite{bes3-D02kenu}.   
In addition, the input parameters $m_0$, $g_1$, and $g_2$~\cite{bes2f980} related to $\Delta\Gamma^{i}_{\rm exp}$
are also considered by varying them within $\pm 1 \sigma$ from their central values.
The largest deviations of the form factor, respectively 2.2\%, 1.2\% and 6.0\%, are taken as systematic uncertainties.
The quadrature sum of the above uncertainties is 6.9\%, which is taken as the total systematic uncertainty. 

\begin{figure}[htp]
\begin{center}
\includegraphics[width=0.485\textwidth]{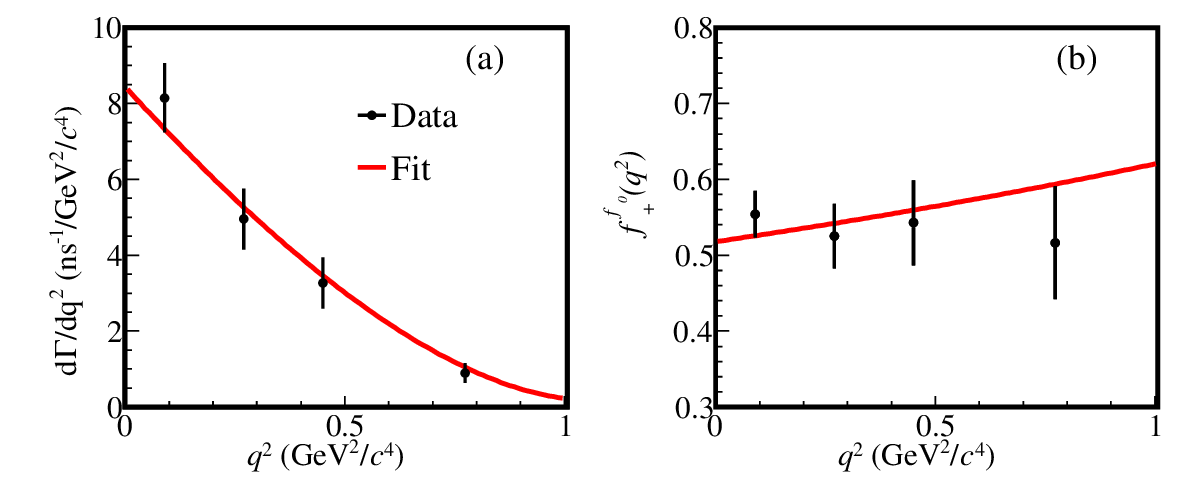}
\put(-185,53){\fontsize{6}{6} { $\chi^2/{\rm NDF}=0.8$} }
\caption{ Fit to the differential decay rate as a function of $q^2$ (a) and projection to the form factor $f^{f_{0}}_{+}(q^2)$ (b). 
The points with error bars are data, and the red line is the fit.}
\label{fit:ff-width}
\end{center}
\end{figure}

The fit to the differential decay rate of the channel $D_{s}^{+}\to f_{0}(980)e^{+}\nu_{e}$ and the form factor projection are shown in Fig.~\ref{fit:ff-width}.
Using the form factor parametrization of Eq.~(\ref{eq:rate}) and the Flatt\'e formula Eq.~(\ref{eq:Ps}) for the $f_0(980)$ decay in the fit,
the product of the form factor and $|V_{cs}|$ is determined to be $f^{f_0}_+(0)|V_{cs}|=0.504\pm0.017_{\rm stat}\pm0.035_{\rm syst}$.
The fit result is shown in Fig.~\ref{fit:ff-width} (a), while Fig.~\ref{fit:ff-width} (b) shows the same fit in projection to the form factor $f^{f_{0}}_{+}(q^2)$.

\papersection{Summary and discussions}
Using $e^+e^-$ collision data corresponding to an integrated luminosity of 
7.33${~\rm fb^{-1}}$ collected at $E_{\rm CM}$ = ${\rm 4.128-4.226~GeV}$ by the BESIII detector, 
we measure the BF of the decay $D^+_s\to f_0(980)e^{+}\nu_{e}$ with $f_0(980)\to\pi^+\pi^-$ to be $(1.72\pm0.13_{\rm stat}\pm0.10_{\rm syst})\times 10^{-3}$,
which is 2.6 times more accurate than the previous measurement~\cite{cleo2015-Ds2semilep}.
In the $q\bar{q}$ mixture picture, this implies that the $s\bar{s}$ component dominates for the $f_0(980)$.
An upper limit on the BF is set to be ${\mathcal B}[D^+_s\to f_0(500) e^+\nu_{e},f_0(500)\to\pi^+\pi^-] < 3.3 \times 10^{-4}$ at 90\% confidence level with estimation of the negligible nonresonance contribution.
Moreover, these BF measurements, especially the one involving the $f_0(500)$, favor the predictions~\cite{RMWang2023,YKHsiao2023} of models assuming the $f_0(980)$ and $f_0(500)$ as tetraquark composition over those based on the $q\bar{q}$ mixture picture.
This is consistent with the arena of $D^+$ semileptonic decays~\cite{bes3-D2f0enu}.

Furthermore, we determine $f^{f_0}_+(0)|V_{cs}|=0.504\pm0.017_{\rm stat}\pm0.035_{\rm syst}$ for the first time
by analyzing the dynamics of $D^+_s\to f_0(980)e^{+}\nu_{e},~f_0(980)\to\pi^+\pi^-$ decay.
Using $|V_{cs}|=0.973~49\pm0.000~16$~\cite{pdg}, we obtain $f^{f_0}_+(0)=0.518\pm0.018_{\rm stat}\pm0.036_{\rm syst}$.
In Table~\ref{comFF} the measured form factor result at $q^2=0$ is compared with different theoretical predictions.
Our measurement agrees with the predictions in Refs.~\cite{qqbar2,QCDSR1,QCDSR2},
but is much higher than the predictions in Refs.~\cite{LCSR1,LF-RQM,Soni2020}.
It is notable that most predictions for the form factor $f^{f_0}_{+}(0)$ and BF depend on the angle $\phi$, which is only known with large uncertainty.
So, the measured form factor and BF are both important to constrain this angle and probe the quark component in $f_0(980)$~\cite{LF-RQM}.
Although most theoretical predictions for $f^{f_0}_{+}(0)$ 
have a large uncertainty due to the $\phi$ uncertainty, the measured form factor line shape is a powerful tool to distinguish different models. 
Finally, these results are important to understand the nature of the light scalar states $f_0(980)$ and $f_0(500)$, and the nonperturbative dynamics of charm meson decays.

The BESIII Collaboration thanks the staff of BEPCII and the IHEP computing center for their strong support. 
The authors are grateful to De-Liang Yao, Shan Cheng and Xian-Wei Kang for valuable discussions.
This work is supported in part by National Key R\&D Program of China under Contracts No. 2020YFA0406400, No. 2023YFA1606000, No. 2020YFA0406300; National Natural Science Foundation of China (NSFC) under Contracts No. 11635010, No. 11735014, No. 11835012, No. 11935015, No. 11935016, No. 11935018, No. 11961141012, No. 12022510, No. 12025502, No. 12035009, No. 12035013, No. 12061131003, No. 12192260, No. 12192261, No. 12192262, No. 12192263, No. 12192264, No. 12192265; 
Natural Science Foundation of Hunan Province, China under Contract No.~2021JJ40036
and the Fundamental Research Funds for the Central Universities under Contract No.~020400/531118010467;
the Chinese Academy of Sciences (CAS) Large-Scale Scientific Facility Program; the CAS Center for Excellence in Particle Physics (CCEPP); Joint Large-Scale Scientific Facility Funds of the NSFC and CAS under Contract No. U1832207; CAS Key Research Program of Frontier Sciences under Contracts No. QYZDJ-SSW-SLH003, No. QYZDJ-SSW-SLH040; 100 Talents Program of CAS; The Institute of Nuclear and Particle Physics (INPAC) and Shanghai Key Laboratory for Particle Physics and Cosmology; ERC under Contract No. 758462; European Union's Horizon 2020 research and innovation programme under Marie Sklodowska-Curie grant agreement under Contract No. 894790; German Research Foundation DFG under Contracts No. 443159800, No. 455635585, Collaborative Research Center CRC 1044, FOR5327, GRK 2149; Istituto Nazionale di Fisica Nucleare, Italy; Ministry of Development of Turkey under Contract No. DPT2006K-120470; National Research Foundation of Korea under Contract No. NRF-2022R1A2C1092335; National Science and Technology Fund of Mongolia; National Science Research and Innovation Fund (NSRF) via the Program Management Unit for Human Resources and Institutional Development, Research and Innovation of Thailand under Contract No. B16F640076; Polish National Science Centre under Contract No. 2019/35/O/ST2/02907; The Royal Society, UK under Contract No. DH160214; The Swedish Research Council; U. S. Department of Energy under Contract No. DE-FG02-05ER41374.

\onecolumngrid

\begin{table}[!htp]
\centering
\caption{Comparison of the form factor at $q^2=0$ between our measurement and various theoretical predictions
(CLFD: covariant light-front dynamics; DR: dispersion relation; QCDSR: QCD sum rule; LCSR: light-cone QCD sum rules; LFQM: light-front quark model; CCQM: covariant confined quark model).
}
\vspace{0.25cm}
\scalebox{0.90}{
\begin{tabular}{c|c|c|c|c|c|c|c|c}
\hline
\hline
                     & This work     &CLFD~\cite{qqbar2}  &DR~\cite{qqbar2}  &QCDSR~\cite{QCDSR1} &QCDSR~\cite{QCDSR2} &LCSR~\cite{LCSR1} &LFQM~\cite{LF-RQM} &CCQM~\cite{Soni2020} \\
\hline
$f^{f_0}_{+}(0)$     & $ 0.518\pm0.018_{\rm stat}\pm0.036_{\rm syst} $ &0.45  &0.46  &$0.50\pm0.13$ & $0.48\pm0.23$ &$0.30\pm0.03$ & $0.24\pm0.05$ & $ 0.36\pm0.02 $ \\
\hline
Difference ($\sigma$)& --- & 1.7  & 1.4  &0.1           &0.2            & 4.3          & 4.3            & 2.8 \\
\hline
$\phi$ in theory     & --- & $(32\pm4.8)^\circ$  & $(41.3\pm5.5)^\circ$ &$35^\circ$ & $(8^{+21}_{-8})^\circ$ & --- &  $(56\pm7)^\circ$ & $31^\circ$ \\
\hline
\hline
\end{tabular}
\label{comFF}
}
\end{table}
\twocolumngrid

\appendix
\subsection{Appendix~A: Introduction to the tag technique }
\label{app:a}
The semileptonic $D_{s}^{+}$ decays can be studied with a tag technique in the process $e^{+}e^{-} \to D_{s}^{*\pm}D_{s}^{\mp} \to \gamma D_{s}^{+}D_{s}^{-}$
where the neutrino is only undetected in the final states.
There are two types of samples used in the tag technique: tag sample and double tag sample.
In the tag sample, the $D_{s}^{-}$ mesons are reconstructed through various hadronic decays.
In the double tag sample appropriately designated as the ``signal'' sample in this Letter, besides the tag $D_{s}^{-}$, the semileptonic signal $D_{s}^{+}$
and the transition photon from the decay $D_{s}^{*\pm}\to \gamma D_{s}^{\pm}$ are reconstructed with the remaining charged tracks and neutral showers. 
The BF with this tag technique is obtained with the following formula:
\begin{equation}
{\cal B}_{\rm sig}=\frac{N^{\rm tot}_{\rm sig}}{{\cal B}_{\gamma}\times\sum_{i} N^{i}_{\rm tag}\times (\epsilon^{i}_{\rm tag,sig}/\epsilon^{i}_{\rm tag})}
=\frac{N^{\rm tot}_{\rm sig}}{ {\cal B}_{\gamma} \times N^{\rm tot}_{\rm tag} \times \bar{\epsilon}},
\label{eq:Bsig-gen}
\end{equation}
where $N^{i}_{\rm tag}$, $\epsilon^{i}_{\rm tag}$ and $\epsilon^{i}_{\rm tag,sig}$ are tag yield,
tag efficiency and signal efficiency with the present tag for the $i$th tag mode, respectively;
${\cal B}_{\gamma}$ is the BF of $D^{*\pm}_s \to\gamma D^{\pm}_s$ decay;
$\bar{\epsilon}=\sum_{i}[(N_{\rm tag}^{i}/{N_{\rm tag}^{\rm tot}})\times(\epsilon_{\rm tag,sig}^{i}/\epsilon_{\rm tag}^{i})]$ is the weighted efficiency.

\subsection{Appendix~B: Systematic uncertainties of the BF measurement }
\label{app:b}
For the decay $D_{s}^{+}\to f_{0}(980) e^{+} \nu_{e}$ with $f_{0}(980)\to\pi^+\pi^-$, there are the below systematic uncertainties of the BF measurement.
The systematic uncertainties of the tracking or particle identification efficiencies of $\pi^{\pm}$ and $e^{+}$ are studied 
with control samples of $e^+e^- \to K^+K^-\pi^{+}\pi^{-}$ and $e^+e^- \to \gamma e^+e^-$ processes.
For $e^{+}$ and $\pi^{\pm}$, both the tracking and particle identification uncertainties are assigned to be 0.5\% and 1.0\% for the $\pi^{+}\pi^{-}$ pair.
The uncertainty from the quoted BF of $D^{*\pm}_s\to\gamma D^{\pm}_s$ decay is 0.7\%~\cite{pdg}.
The uncertainty due to the transition photon reconstruction is estimated to be 2.0\% using
the control sample of $e^{+}e^{-}\to D_{s}^{*\pm}D_{s}^{\mp}$ events, where $D_{s}^{-}$ decays via a tag mode,
while $D_{s}^{+}$ decays via one of the two hadronic channels: $D_{s}^{+}\to K_{S}^{0}K^{+}$ or $D_{s}^{+}\to K^{+}K^{-}\pi^{+}$.
The uncertainty in the total number of the tag $D_{s}^{-}$ mesons is assigned to be 0.3\% by 
examining the changes of the fit yields when varying the signal shape, background shape, and taking into account the background fluctuation in the fit.
The uncertainty associated with the signal MC model is estimated to be 4.4\% by replacing the $f_0(980)$ line shape
from BESII~\cite{bes2f980} with the one from LHCb~\cite{LHCbf980} in generating the signal MC samples.
The uncertainty of the $M_{\pi^+\pi^-}$ fit is estimated to be 2.1\% by altering the nominal MC background shape.
Firstly, we use alternative MC shapes where the relative fractions of backgrounds from continuum and non-$D_{s}^{*\pm} D_{s}^{\mp}$ open-charm processes are varied by $\pm30$\%
according to the uncertainties of their assigned cross sections in the inclusive MC sample.
Secondly, we vary the relative fraction ($\pm 1\sigma$) of the peaking background channel $D^+_s\to\eta^{\prime}(\gamma\pi^+\pi^-)e^+\nu_e$~\cite{pdg}.
The uncertainty due to neglecting other $S$-wave contributions was found to be negligible.
The total systematic uncertainty is 5.6\%, obtained by adding all contributions in quadrature.

For the decay $D_{s}^{+}\to f_{0}(500) e^{+} \nu_{e}$ with $f_{0}(500)\to\pi^+\pi^-$, 
there are two types of systematic uncertainties of the upper limit on the BF measurement: additive and multiplicative.
The additive uncertainty is dominated by the background shape.
Besides the same uncertainty sources following the previous BF measurement,
we change the relative fraction~($\pm 1\sigma$) of the major background of $D_{s}^{+} \to \eta e^{+} \nu_{e}$ decay~\cite{pdg}
and the peak background of $D_{s}^{+}\to f_{0}(980) e^{+} \nu_{e}$ decay.
The multiplicative uncertainties are the same as those for $D_{s}^{+}\to f_{0}(980) e^{+} \nu_{e}$ decay except for the signal MC model efficiency.
This uncertainty is estimated to be 1.1\% by varying the parameters of the Bugg line shape~\cite{bugg} within their uncertainties in generating the signal MC samples.

\onecolumngrid
\section*{}
\begingroup
\small
\renewcommand{\author}[1]{\begin{center}#1\end{center}}
\author{
M.~Ablikim$^{1}$, M.~N.~Achasov$^{13,b}$, P.~Adlarson$^{75}$, R.~Aliberti$^{36}$, A.~Amoroso$^{74a,74c}$, M.~R.~An$^{40}$, Q.~An$^{71,58}$, Y.~Bai$^{57}$, O.~Bakina$^{37}$, I.~Balossino$^{30a}$
% ...
\\
\vspace{0.2cm}
(BESIII Collaboration)\\
\vspace{0.2cm} {\it
$^{1}$ Institute of High Energy Physics, Beijing 100049, People's Republic of China\\
% ...
$^{13}$ G.I. Budker Institute of Nuclear Physics SB RAS (BINP), Novosibirsk 630090, Russia\\
% ...
$^{30a}$ INFN Sezione di Ferrara, INFN Sezione di Ferrara, I-44122, Ferrara, Italy\\
$^{30b}$ INFN Sezione di Ferrara, University of Ferrara,  I-44122, Ferrara, Italy\\
% ...
$^{36}$ Johannes Gutenberg University of Mainz, Johann-Joachim-Becher-Weg 45, D-55099 Mainz, Germany\\
$^{37}$ Joint Institute for Nuclear Research, 141980 Dubna, Moscow region, Russia\\
% ...
$^{40}$ Liaoning Normal University, Dalian 116029, People's Republic of China\\
% ...
$^{57}$ Southeast University, Nanjing 211100, People's Republic of China\\
$^{58}$ State Key Laboratory of Particle Detection and Electronics, Beijing 100049, Hefei 230026, People's Republic of China\\
% ...
$^{71}$ University of Science and Technology of China, Hefei 230026, People's Republic of China\\
% ...
$^{74a}$ University of Turin and INFN, University of Turin, I-10125, Turin, Italy\\
$^{74b}$ University of Turin and INFN, University of Eastern Piedmont, I-15121, Alessandria, Italy\\
$^{74c}$ University of Turin and INFN, INFN, I-10125, Turin, Italy\\
$^{75}$ Uppsala University, Box 516, SE-75120 Uppsala, Sweden\\
% ...
$^{b}$ Also at the Novosibirsk State University, Novosibirsk, 630090, Russia\\
% ...
}
}

\endgroup
\twocolumngrid
\maketitle

\end{document}